\documentclass[11pt]{article}
\usepackage{graphicx}
  \usepackage{amsthm,amsfonts}
  \usepackage{amsmath}
\newcommand{\bea}   {\begin{eqnarray}}
\newcommand{\eea}   {\end{eqnarray}}

\begin{document}
\renewcommand{\thefootnote}{\fnsymbol{footnote}}

\thispagestyle{empty}

\title{A world-line framework for $1D$ Topological Conformal $\sigma$-models}

\author{L. Baulieu\thanks{{E-mail: {\em baulieu@lpthe.jussieu.fr}}},\quad 
N. L. Holanda\thanks{{E-mail: {\em linneu@cbpf.br}}}
\quad and\quad F.
Toppan\thanks{{E-mail: {\em toppan@cbpf.br}}}
\\
\\
}
\maketitle

\centerline{$^{\ast}$
{\it Sorbonne Universit\'es, LPTHE, Universit\'e Pierre et Marie Curie,}}
{\centerline {\it\quad
CNRS-UMR 7589, 4 place Jussieu, F-75252 Paris, France.}}
\centerline{$^{\dag}$ $^{\ddag}$
{\it CBPF, Rua Dr. Xavier Sigaud 150, Urca,}}{\centerline {\it\quad
cep 22290-180, Rio de Janeiro (RJ), Brazil.}
~\\
\maketitle
\begin{abstract}
We use world-line methods for pseudo-supersymmetry to construct $sl(2|1)$-invariant actions for the 
$(2,2,0)$ chiral and ($1,2,1)$ real supermultiplets  of the twisted $D$-module representations of the $sl(2|1)$ superalgebra. The derived one-dimensional topological conformal $\sigma$-models are invariant under nilpotent operators. The actions are constructed for both parabolic and hyperbolic/trigonometric realizations (with extra potential terms in the latter case). The scaling dimension $\lambda$ of the supermultiplets defines a coupling constant, $2\lambda+1$, the free theories being recovered at $\lambda=-\frac{1}{2}$.
We also present, generalizing previous works, the $D$-module representations of one-dimensional superconformal algebras induced by ${\cal N}=(p,q)$ pseudo-supersymmetry acting on $(k,n,n-k)$ supermultiplets. Besides $sl(2|1)$,
we obtain the superalgebras $A(1,1)$, $D(2,1;\alpha)$, $D(3,1)$, $D(4,1)$, $A(2,1)$ from $(p,q)= (1,1), (2,2), (3,3), (4,4), (5,1)$, at given $k,n$ and critical values of $\lambda$. \par
~\\\end{abstract}
\vfill

\rightline{CBPF-NF-004/15
}

\newpage
\section{Introduction}
In this paper, using constraints from world-line pseudo-supersymmetry and some of the results in \cite{pap} and  \cite{hoto}, we generalize the class of one-dimensional $sl(2|1)$-invariant conformal topological $\sigma$-models of \cite{bato2}.
Specifically, we construct invariant actions for any real value $\lambda$ of the scaling dimension, for both chiral and real supermultiplets and in both  parabolic and hyperbolic/trigonometric cases of the $D$-module representations of the $sl(2|1)$ superalgebra. For $\lambda\neq -\frac{1}{2}$ the invariant actions possess an interacting term (only the free actions were discussed in \cite{bato2}). \par
A further result of this paper is the extension of the criticality condition of the $D$-module representations of one-dimensional superconformal algebras from the case of positive supersymmetry (analyzed in \cite{{kuto},{khto},{hoto}}) to the case of pseudo-supersymmetry ${\cal N}=(p,q)$ with both $p,q>0$.
The extension to pseudo-supersymmetry has non-trivial consequences. Unlike the positive supersymmetry, criticality of $D$-module representation is already encountered for supermultiplets with two bosonic and two fermionic component fields, the $D(2,1;\alpha)$ superalgebras being induced by ${\cal N}=(2,2)$ acting on the chiral supermultiplet. Furthermore, free parameters enter the construction of topological charges, which are nilpotent operators obtained from linear combinations of the pseudo-supersymmetry generators. We will see in which cases these parameters have to be fixed in order to guarantee the existence of superconformally invariant actions. Some other interesting features appear w.r.t. the positive superconformal case. The boost operator $K$ (the $sl(2)$ conformal partner of the Hamiltonian $H$), for instance, can be non-diagonal (a non-diagonal $K$ operator is required by superconformal invariance of the real supermultiplet actions).\par
The framework of our paper is that of (one-dimensional) topological quantum field theory, realized as an $N=2$ supersymmetric quantum mechanics with twisted superalgebra, see \cite{{bagr},{brt},{basi},{bato}}.
Different from our worldline approach, a construction based on target manifold restrictions \cite{{alfr},{wit2},{wit3},{copa},{fks}} has been discussed for $(n,n)$ superconformal mechanics in \cite{sin}.
\par
There is a vast literature on (positive, i.e. untwisted) superconformal mechanics \cite{{akpa},{fura}}, its model building and its applications (which range from $AdS_2/CFT_1$ correspondence, near-horizon geometry of extremal black holes, etc.).  One can consult the two review papers \cite{{bmsv},{fil1}} and the references therein.\par
It is useful to present our terminology and notations.  \par
The ${\cal N}=(p,q)$ pseudo-supersymmetry algebra is given by $p+q$ fermionic operators ${\text Q}_I$ and a single bosonic operator $H$ (the Hamiltonian), as 
\bea\label{pseudo}
\{{\text Q}_I,{\text Q}_J \}= \eta_{IJ} H, && [H,{\text Q}_I]=0,
\eea
where $p$ entries of the diagonal matrix $\eta_{IJ}$ are $+1$ and $q$ entries are $-1$ (we write, accordingly, $\{{\text Q}_I\}\equiv\{ Q_i^+, Q_j^-\}$ for $i=1,\ldots , p$, $j=1,\ldots q$, so that ${Q_\star^\pm}^2=\pm H$).  For $q=0$ we recover the ($p$-extended) superalgebra of the Supersymmetric Quantum Mechanics \cite{wit}. The $q=0$ case is called the ``positive" or ``untwisted"  supersymmetry. \par The $D$-module representations of (\ref{pseudo}), in terms of differential operators in one variable (the ``time" coordinate) have been constructed in \cite{{pato},{kuroto}}. They act on supermultiplets of time-dependent component fields with different scaling dimensions. For the case of interest here the supermultiplets are of the form $(k,n,n-k)$, namely possessing $k$ bosons of scaling dimension $\lambda$, $n$ fermions of scaling dimension $\lambda+\frac{1}{2}$, $n-k$ bosons of scaling dimension $\lambda+1$. As customary, with a slight abuse of language (this is only true when invariant actions are constructed), the first set of bosons will be referred to as the ``propagating bosons", the remaining ones are referred to as the ``auxiliary fields".\par
For ${\cal N}=(2,2)$ the irreducible supermultiplets are $(2,2,0)$ (also known as the ``chiral supermultiplet") and
$(1,2,1)$ (also known as the ``real supermultiplet").\par
The finite one-dimensional superconformal algebras are the simple Lie superalgebras \cite{{kac},{fss}} ${\cal G}$ which can be decomposed according to the grading
${\cal G} = {\cal G}_{-1}\oplus {\cal G}_{-\frac{1}{2}}\oplus
{\cal G}_0\oplus {\cal G}_{\frac{1}{2}}\oplus {\cal G}_{1}$
and whose even sector is ${\cal G}_{even}=sl(2)\oplus R$ (the subalgebra $R$ is known as $R$-symmetry). The odd sector (${\cal G}_{\frac{1}{2}}\oplus {\cal G}_{-\frac{1}{2}}$) is spanned by $2m$ generators.
The positive sector ${\cal G}_{>0}$ is isomorphic to the (\ref{pseudo}) superalgebra for $m=p+q$. If $D,H,K$ are the $sl(2)$ generators, with $D$ being the Cartan element, ${\cal G}_1$ (${\cal G}_{-1}$) is spanned by $H$ ($K$), while ${\cal G}_0=D{\mathbb C}\oplus R$. 
\par
Since (\ref{pseudo}) is the ${\cal G}_{>0}$ subalgebra, under certain (critical) conditions, the $D$-module representations
of (\ref{pseudo}) can be extended to a differential realization for an associated superconformal algebra, see \cite{{kuto},{khto},{hoto}} for the $q=0$ case.
Following \cite{hoto}, the ``parabolic" $D$-module reps of \cite{{kuto},{khto}} can be mapped into ``hyperbolic/trigonometric" $D$-module reps via a similarity transformation and a change of the time coordinate. These $D$-module reps are written in terms of hyperbolic/trigonometric functions (hence, a dimensional parameter $\mu$ is introduced). The presence of $\mu$ allows for extra potential terms (absent in the parabolic case)  in the invariant actions \cite{{pap},{hoto}}. What characterizes the parabolic versus the hyperbolic/trigonometric $D$-module reps is the fact that, in the latter case, the Cartan generator $D$ is proportional to the time-derivative operator, while in the parabolic case the time-derivative operator is proportional to $H$, the $sl(2)$ positive root. Since trigonometric $D$-module reps are recovered from the hyperbolic $D$-module reps via an analityc continuation of $\mu$ \cite{hoto}, from now on, for simplicity,
we just call ``hyperbolic" this class of $D$-module reps.\par
The twisted supersymmetry is the presentation of a superalgebra in terms of nilpotent fermionic generators.
Following \cite{bato} a nilpotent operator $Q$ can be constructed as a linear combination of the
pseudo-supersymmetry generators ${\text Q}_I$ in (\ref{pseudo}) as $Q=\sum_I c_I {\text Q}_I$. The coefficients $c_I$ have to be chosen to guarantee $Q^2=0$. All $c_I$'s have to be real to respect the reality condition of the superalgebra. Therefore, the nilpotency of $Q$ requires a pseudo-supersymmetry with both $p,q>0$. In the case of the $sl(2|1)$ superalgebra its twisted version is expressed with two nilpotent operators $Q_1, Q_2$ ($Q_1^2=Q_2^2=0$, $\{Q_1,Q_2\}=H$) and their two superconformal partners. \par
In the functional integral approach, via path-integral, to twisted supersymmetry, the passage from a $2n$ positive supersymmetry to a $(n,n)$ pseudo-supersymmetry is formally realized via complex transformations. When dealing, as here, with representation theory, this passage is not allowed because we need to respect the real form of the superalgebras. The connection between twisted and untwisted versions of  supersymmetric theories has been discussed in \cite{bato}. 
Getting nilpotent supersymmetry generators  provides a good viewpoint for investigating the properties of  supersymmetry since many questions can be reduced to cohomological questions, with  $Q$-invariance often realized as $Q$-exactness. Depending on which problem one considers,  various interesting properties of supersymmetric theories can be better understood either in the twisted or in the untwisted formulation.
\par
Starting from the (\ref{pseudo}) pseudo-supersymmetry operators acting on a given supermultiplet,
the following steps have to be fulfilled in order to obtain a one-dimensional, topological, conformally invariant, $\sigma$-model:\\
{\em i}) at first a set of nilpotent operators $Q_i$'s has to be chosen by properly selecting, for any given $i$, the $c_I$'s coefficients (a certain arbitrariness can be reflected in the appearance of free parameters);\\
{\em ii}) next, the most general boost operator $K$  which, together with the $Q_i$'s operators, allow to construct a superconformal algebra, must be determined;
\\
{\em iii}) finally, the superconformal algebra is implemented as a symmetry by constructing an action manifestly invariant under the $Q_i$'s transformations (for positive supersymmetry this procedure is discussed in \cite{kuroto}) and by imposing (following \cite{{khto},{kuto},{hoto}}), the invariance under the $K$ transformation as a constraint.   
\par
These three steps are presented in Sections {\bf 2} and {\bf 3}.
We postpone to the Conclusions the discussion of the results here obtained.\par

The scheme of the paper is as follows. In Section {\bf 2} we discuss the $D$-module representations of the $sl(2|1)$ superconformal algebra realized on supermultiplets with two bosons and two fermions. Even if it is not the largest superalgebra acting on these supermultiplets, $sl(2|1)$ is the largest symmetry superalgebra of the invariant actions. We present in Section {\bf 3} the construction of the $sl(2|1)$-invariant $\sigma$-models.
In Section {\bf 4} we extend the results of \cite{{kuto}, {khto}, {hoto}}, presenting the list of $D$-module reps of superconformal algebras induced by the supermultiplets associated with the pseudo-supersymmetry. In Appendix {\bf A} the smallest topological conformal algebra and its invariant actions are introduced. In Appendix {\bf B} the $D$-module reps discussed in Section {\bf 4} are explicitly constructed. In the Conclusions we comment our results and point out the directions of future development.

\section{Symmetry algebra of the topological conformal $\sigma$-models}

For the models under consideration the superconformal symmetry algebra contains in the bosonic sector $4$ generators,
$H$, $D$, $K$ and $G$, which close the $sl(2)\oplus u(1)$ algebra. Its non-vanishing commutators are
\begin{equation}
\label{al1}
[H,D] = H,\quad [K,D] = -K,\quad [H,K] = 2D.
\end{equation}
$H,D,K$ generates the one-dimensional conformal algebra $sl(2)$, while the $u(1)$ charge $G$ is the $R$-symmetry operator.\par
The topological conformal extension of $sl(2)\oplus u(1)$ requires the introduction of $4$ extra nilpotent fermionic generators
(the twisted supersymmetry operators $  Q_1$, $  Q_2$ and their superconformal partners  $\bar Q_1$ and $\bar Q_2$, so that $Q_1^2= Q_2^2=\bar{Q}_1^2= \bar{Q}_2^2=0$). The consistency condition, provided by the graded-Jacobi identities \cite{fss}, determines the structure constants of the superalgebra. The extra non-vanishing (anti)commutators are
\begin{equation}
\label{al2}
\{Q_k,Q_j\} = (\sigma_1)_{kj}H ,\quad \{Q_k,\bar{Q}_j\} = -(\sigma_1)_{kj}D+i(\sigma_2)_{kj}G ,\quad
 \{\bar{Q}_k,\bar{Q}_j\} = (\sigma_1)_{kj}K, 
\end{equation}
together with
\begin{equation}
\label{al3}
[K,Q_k] = \bar{Q}_k
, \quad [H,\bar{Q}_k] = -Q_k,\quad \nonumber
\end{equation}
\begin{equation}
\label{eq3}
[Q_k,D] = \frac{Q_k}{2},\quad [\bar{Q}_k,D] = -\frac{\bar{Q}_k}{2},\nonumber
\end{equation}
\begin{equation}
\label{eq4}
[Q_k, G] = (\sigma_3)_{kj}\frac{Q_j}{2},\quad  [\bar{Q}_k, G] = (\sigma_3)_{kj}\frac{\bar{Q}_j}{2}.
\end{equation}
In the r.h.s. the Pauli matrices {$
\sigma_1 = {\tiny \left(\begin{array}{cc}
0 & 1\\
1 & 0
\end{array}\right) }, \quad
\sigma_2 = {\tiny \left(\begin{array}{cc}
0 & -i\\
i & 0
\end{array}\right)},  \quad \sigma_3 = {\tiny \left(\begin{array}{cc}
1 & 0\\
0 & -1
\end{array}\right)}
$
}
have been used. One should note that the structure constants of the superalgebra are real.\par

The commutator with $D$ defines  the standard conformal weights: $+1$ ($H$), $0$ ($D,G$), $-1$ ($K$) and
  $+\frac{1}{2} $ ($Q_1,Q_2$), $-\frac{1}{2} $ ($\bar Q_1, \bar Q_2$).


Interestingly enough, the topological superalgebra  also possesses   a conserved bigrading, defined for each operator  as follows
\begin{equation*}
H^{1,-1},\quad D^{0,0}, \quad G^{0,0},\quad  K^{-1,1},\quad   Q_1^{1,0},\quad  Q_2^{0,-1},\quad    \bar  Q_1^{-1,0},\quad  \bar Q_2^{0,1}.\quad 
\end{equation*}
The sum of both values of the indices of the  bigrading can be called the {\em ghost number}. Operators and fields with even (odd) ghost number are commuting (anticommuting). If the bigrading is enforced, the operation of {\em untwisting} the (\ref{al1}-\ref{al3}) algebra (namely, presenting the fermionic generators in a pseudo-supersymmetry diagonal basis) is not allowed, because linear combinations of the fermionic generators do not preserve the bigrading. On the other hand, the untwisting operation is allowed if we just impose the conservation of the ghost number (modulo two). The duality relating $H$ and $K$, $Q_k$ and $\bar Q_k$, corresponds to a parity transformation which
preserves the bigrading.\par
One should note that, once the conformal boost operator  $K$ and the $Q_k$'s operators have been determined  in a given representation, the conformal partners   $\bar{Q}_k$'s  are obtained from the relation $[K,Q_k] = \bar{Q}_k$. 
 

When computing, in Section {\bf 3}, invariant actions we assume for the Lagrangians the vanishing of the ghost number. Other invariants (such as anomalies and cocycles of higher degrees), possessing different values of the  ghost number, exist. These invariants can be called the ``topological observables".\par

The topological conformal algebra
(\ref{al1}-\ref{al3}) is a real form of the $A(1,0)\approx sl(2|1)$ superalgebra. $A(1,0)$ is contained as a subalgebra in the superalgebras $D(2,1;\alpha)$ and $A(1,1)$ with its central extension. In appendix {\bf B} it is shown that the irreducible $D$-module representations of these superalgebras are realized on supermultiplets with $2$ bosonic and $2$ fermionic component fields. Clearly, on the same set of fields, a $D$-module rep of $A(1,0)$ is induced. The key point is that $A(1,0)$ is the largest subalgebra which can emerge as a symmetry algebra of a topological action constructed with this basic set of $2+2$ component fields.\par
In this Section we pave the way for the construction of the topological conformal $\sigma$-models of Section {\bf 3} by discussing the properties of the induced $A(1,0)$ $D$-module representations. As mentioned in the Introduction, the representations can be either parabolic or hyperbolic; the basic supermultiplets in this case are either $(2,2,0)$ (the chiral supermultiplet) or $(1,2,1)$ (the real one).

\subsection{The $(2,2,0)$ $D$-module reps of the $sl(2|1)$ superalgebra}

The operators are assumed to act upon the $(2,2,0)$ supermultiplet $(x,\bar{x};\psi,\bar{\psi})$, with $x$, $\bar{x}$ propagating bosons. In the parabolic case the representation is constructed as follows. One starts with the four fermionic operators, see Appendix {\bf B}, $Q_1^\pm$, $Q_2^\pm$ of the ${\cal N}=(2,2)$ superconformal algebra $(2,2,0)$ $D$-module rep. Explicitly,
{\tiny\begin{eqnarray}
\label{eq7}
&Q_1^+ = \left(\begin{array}{cccc}
0 & 0 & 0 & 1 \\
0 & 0 & 1 & 0\\
0 & \partial_t & 0 & 0\\
\partial_t & 0 & 0 & 0 \\
\end{array}\right),\quad\quad
Q_1^- = \left(\begin{array}{ccccc}
0 & 0 & 0 & 1 \\
0 & 0 & -1 & 0\\
0 & \partial_t & 0 & 0\\
-\partial_t & 0 & 0 & 0 \\
\end{array}\right),&\nonumber\\
&Q_2^+ = \left(\begin{array}{cccc}
0 & 0 & 1 & 0 \\
0 & 0 & 0 & -1\\
\partial_t & 0 & 0 & 0\\
0 & -\partial_t & 0 & 0 \\
\end{array}\right),\quad
Q_2^- = \left(\begin{array}{cccc}
0 & 0 & 1 & 0 \\
0 & 0 & 0 & 1\\
-\partial_t & 0 & 0 & 0\\
0 & -\partial_t & 0 & 0 \\
\end{array}\right).&\nonumber
\end{eqnarray} 
} 
They are the square roots of the $\pm H$ operator.  The following two nilpotent operators
can be constructed for an arbitrary real value of the parameter $\beta$,
\bea\label{qbeta}
Q_1 = \frac{Q_1^+ + Q_1^-}{2},\quad &&  \quad
Q_2^{(\beta)} = \frac{Q_1^+ - Q_1^-}{2} + \beta\frac{Q_2^+ + Q_2^-}{2}.
\eea
The next step is considering the most general realization of the $sl(2)$ algebra which is compatible with the scaling dimension of the fields.  
Let ${\mathbb{ I}}_4$ be the $4\times 4$ identity matrix and $\Lambda$ the matrix $\Lambda = diag(\lambda, \lambda,\lambda+\frac{1}{2}, \lambda+\frac{1}{2})$, expressed in terms of the real scaling dimension parameter  $\lambda$. We can assume without loss of generality that the Borel generators $H,D$ are realized in a diagonal form:
\bea
\label{HDpar}
H ={\mathbb{I}}_{4}\cdot \partial_t, \quad &&\quad D={\mathbb{ I}}_{4 }\cdot t\partial_t + \Lambda.
\eea
The most general form of the operator $K$ which preserves the dimensionality of the fields is {\tiny
\begin{equation*}
K = \left(\begin{array}{cccc}
t^2\partial_t+2\lambda t & a_{12}t^2\partial_t + b_{12}t &  0 & 0\\
a_{21}t^2\partial_t + b_{21}t & t^2\partial_t + 2\lambda t & 0 & 0\\
0 & 0 &  t^2\partial_t + (2\lambda+1)t& a_{34}t^2\partial_t + b_{34}t\\
0 & 0 & a_{43}t^2\partial_t + b_{43}t & t^2\partial_t + (2\lambda+1)t
\end{array}\right).
\end{equation*}
}
The dimensionless parameters $a_{ij}$'s and $b_{ij}$'s are allowed by power counting. Nevertheless, the closure of the $sl(2)$ algebra, and particularly the condition $[H,K]=2D$, forces us to set all the parameters $a_{ij}$'s, $b_{ij}$'s equal to zero.\par
Therefore, the $(2,2,0)$ $D$-module rep requires a diagonal $K$, given by
\begin{equation}
\label{Kpar}
K ={\mathbb{ I}}_4\cdot t^2\partial_t + 2\Lambda t.
\end{equation}
The two extra fermionic operators ${\bar Q}_1, {\bar Q}_2$ are obtained from the commutators 
${\bar Q}_1= [K, Q_1]$, ${\bar Q}_2= [K, Q_2]$, while the $R$-symmetry generator $G$ is recovered from, e.g., the anticommutator $\{Q_1,{\bar Q}_2\}$, see (\ref{al2}) and (\ref{al3}). It is a straightforward exercise to verify, for any real $\lambda$, the closure of the $sl(2|1)$ superalgebra on this set of operators.\par
Following \cite{pap} and \cite{hoto} a hyperbolic $D$-module representation can be constructed by
\\
{\em i}) performing a similarity transformation on the operators of the parabolic $D$-module rep and\\
{\em ii}) performing a change of the time coordinate.\\
What characterizes the hyperbolic rep with respect to the parabolic rep is the fact that in this case
the time-derivative operator is $H$, the positive root of $sl(2)$. In the hyperbolic case the time-derivative operator (for the new time coordinate) is $D$, the Cartan generator of $sl(2)$ (as a consequence, the energy spectrum of the invariant theories is continuous in the parabolic case and discrete in the hyperbolic case).\par

We present here a set of operators which allows to reconstruct the full set of  $(2,2,0)$ $D$-module rep generators in the hyperbolic case. For arbitrary $\lambda$ and at the special value $\beta=0$ (in Section {\bf 3} we prove that this is the unique value of $\beta$ which produces $sl(2|1)$-invariant actions for the $(2,2,0)$ supermultiplet) we have
\begin{equation}
\label{eq18}
Q_1 = e^{-\frac{\mu t}{2}}\left(\begin{array}{ccccc}
0 & 0 & 0 & 1\\
0 & 0 & 0 & 0\\
0 & \frac{1}{\mu}\partial_t - \lambda & 0 & 0\\
0 & 0 & 0 & 0
\end{array}\right),\quad 
Q_2 = e^{-\frac{\mu t}{2}}\left(\begin{array}{ccccc}
0 & 0 & 0 & 0\\
0 & 0 & 1 & 0\\
0 & 0 & 0 & 0\\
\frac{1}{\mu}\partial_t - \lambda & 0 & 0 & 0
\end{array}\right).
\end{equation}
The $sl(2)$ algebra operators are 
\begin{equation}
\label{eq19}
H = e^{-\mu t}(\mathbb{I}_{4}\cdot \frac{1}{\mu}\partial_t - \Lambda), \quad D=\mathbb{I}_{4}\cdot \frac{1}{\mu}\partial_t,\quad K = e^{\mu t}(\mathbb{I}_{4}\cdot\frac{1}{\mu}\partial_t + \Lambda).
\end{equation}
Here again $\Lambda = diag(\lambda, \lambda,\lambda+\frac{1}{2}, \lambda+\frac{1}{2})$. The remaining operators are obtained from the (anti)commutators in (\ref{al2},\ref{al3}). The parameter $\mu$ is a dimensional parameter which can be set equal to $1$ without loss of generality.

\subsection{The $(1,2,1)$ $D$-module reps of the $sl(2|1)$ superalgebra}

The construction is similar to the previous case. 
The operators now act on the supermultiplet $(x,b;\psi,\bar{\psi})$, where $x$ is a propagating boson and $b$ an auxiliary field. In the parabolic case we start with the four fermionic operators given in Appendix {\bf B},
{\tiny
\begin{eqnarray}
\label{eq22}
&
Q_1^+ = \left(\begin{array}{ccccc}
0 & 0 & 0 & 1 \\
0 & 0 & \partial_t & 0\\
0 & 1 & 0 & 0\\
\partial_t & 0 & 0 & 0 \\
\end{array}\right),\quad
Q_1^- = \left(\begin{array}{ccccc}
0 & 0 & 0 & 1 \\
0 & 0 & -\partial_t & 0\\
0 & 1 & 0 & 0\\
-\partial_t & 0 & 0 & 0 \\
\end{array}\right),&\nonumber\\
&Q_2^+ = \left(\begin{array}{ccccc}
0 & 0 & 1 & 0 \\
0 & 0 & 0 & -\partial_t\\
\partial_t & 0 & 0 & 0\\
0 & -1 & 0 & 0 \\
\end{array}\right),\quad
Q_2^- = \left(\begin{array}{ccccc}
0 & 0 & 1 & 0 \\
0 & 0 & 0 & \partial_t\\
-\partial_t & 0 & 0 & 0\\
0 & -1 & 0 & 0 \\
\end{array}\right).\nonumber
&
\end{eqnarray} 
}
These are the square roots of $\pm H$ in the $(1,2,1)$ $D$-module rep. Two nilpotent operators (the second one depending on an arbitrary parameter $\gamma$) are constructed as follows
\begin{equation}
\label{qgamma}
Q_1 = \frac{Q_1^+ + Q_1^-}{2} , \quad  \quad  Q_2^{(\gamma)} = \frac{Q_1^+ - Q_1^-}{2} - \gamma\frac{Q_2^+ + Q_2^-}{2} . 
\end{equation}
As in the previous case and without loss of generality the $H,D$ generators of the $sl(2)$ Borel subalgebra can be assumed to be diagonal
\begin{equation}
\label{eq24}
H = {\mathbb I}_4\cdot \partial_t,\quad  D = {\mathbb I}_4\cdot t\partial_t + \Lambda.
\end{equation}
The matrix $\Lambda $ is now  $\Lambda = diag(\lambda, \lambda + 1, \lambda + \frac{1}{2}, \lambda + \frac{1}{2})$.\par
The most general form of the operator $K$, preserving the dimensionality of the fields, is now  {\tiny
\begin{equation}
\label{eq25}
K = \left(\begin{array}{cccc}
t^2\partial_t + 2\lambda t & a_{12}t^3\partial_t + b_{12}t^2  & 0 & 0\\
a_{21}t\partial_t + b_{21} & t^2\partial_t + 2(\lambda+1) t & 0 & 0\\
0 & 0  & t^2\partial_t + (2\lambda+1)t& a_{34}t^2\partial_t + b_{34}t\\
0 & 0 & a_{43}t^2\partial_t + b_{43}t & t^2\partial_t + (2\lambda+1)t
\end{array}\right).\nonumber
\end{equation} 
}
Contrary to the previous case, the closure of the $sl(2)$ algebra allows a non-diagonal form for the operator $K$. Indeed, its most general solution is
\begin{equation}
\label{Kdelta}
K = {\mathbb I}_4\cdot t^2\partial_t + 2\Lambda t + \delta E_{21},
\end{equation}
where $E_{21}$ denotes the matrix whose only non-vanishing entry is $1$ in the $(2,1)$ position. The parameter $\delta$ is arbitrary. At $\delta=0$ we obtain a diagonal form for $K$.\par
 As in the previous case, the remaining generators are computed from the (anti)commutators given in
(\ref{al2},\ref{al3}). The $(1,2,1)$ $D$-module rep of the $sl(2|1)$ superalgebra depends on three arbitrary real parameters,
the scaling dimension $\lambda$, the parameter $\gamma$ entering (\ref{qgamma}) and the parameter $\delta$ entering (\ref{Kdelta}). In Section {\bf 3} we prove that the $sl(2|1)$-invariant actions require $\delta\neq 0$ (therefore, a non-diagonal form of $K$). In the invariant actions the parameter $\gamma$ can be rescaled so that, without loss of generality, one can set $\gamma=1$. In its turn $\delta$ has to be fixed at $\delta=-\lambda$.\par
For these special values, the operators which induce the hyperbolic $(1,2,1)$ $D$-module rep of $sl(2|1)$ are 
the following. The two nilpotent operators $Q_1, Q_2$ are
\begin{equation}
\label{eq35}
Q_1 = e^{-\frac{\mu t}{2}}\left(\begin{array}{ccccc}
0 & 0 & 0 & 1\\
0 & 0 & 0 & 0\\
0 & 1 & 0 & 0\\
0 & 0 & 0 & 0
\end{array}\right),\quad
Q_2 = e^{-\frac{\mu t}{2}}\left(\begin{array}{ccccc}
0 & 0 & -1 & 0\\
0 & 0 & \frac{1}{\mu}\partial_t - \lambda-\frac{1}{2} & 0\\
0 & 0 & 0 & 0\\
\frac{1}{\mu}\partial_t  - \lambda & 1 & 0 & 0
\end{array}\right).
\end{equation}
The $sl(2)$ algebra generators are
\begin{equation}
\label{eq36}
H = e^{-\mu t}({\mathbb I}_4\cdot\frac{1}{\mu}\partial_t - \Lambda),\quad  D = {\mathbb I}_4\cdot\frac{1}{\mu}\partial_t,\quad  K = e^{\mu t}({\mathbb I}_4\cdot\frac{1}{\mu}\partial_t + \Lambda -\lambda E_{21}),
\end{equation}
with $\Lambda = diag(\lambda, \lambda+1, \lambda+ \frac{1}{2}, \lambda+ \frac{1}{2})$. 
The remaining $sl(2|1)$ generators are derived from  the (anti)commutators
(\ref{al2},\ref{al3}).

\section{One-dimensional topological conformal $\sigma$-models}

We present here the conformal topological actions which are $sl(2|1)$-invariant. We require global (twisted) supersymmetry and impose the conformal invariance as a constraint. The procedure repeats the one adopted in \cite{hoto} and \cite{bato2} for deriving one-dimensional superconformal $\sigma$-models from the supermultiplets of the untwisted supersymmetry. Contrary to the untwisted case, besides the scaling dimension $\lambda$, extra free parameters enter and label the twisted superconformal representations. As shown in Section {\bf 2}, one extra parameter, $\beta$, enters the $(2,2,0)$ $D$-module representation. Two extra parameters, $\gamma$ and $\delta$, enter the $(1,2,1)$
$D$-module representation. The requirement of the existence of superconformally invariant actions implies assigning a fixed value to the parameters $\beta$ and $\delta$ while, without loss of generality, $\gamma$ can be rescaled to $1$.

\subsection{Superconformal models for the $(2,2,0)$ supermultiplet}

By construction, a global twisted supersymmetry invariance is obtained in terms of the Lagrangian   
\begin{equation*}
\mathcal{L} = Q_1Q_2^{(\beta)}(F\bar{\psi}\psi)
\end{equation*}
where $F= F(x,\bar{x})$ and $Q_1, Q_2^{(\beta)}$ are defined in (\ref{qbeta}). This yields, up to a total time derivative,
\begin{equation}
\label{eq11}
\mathcal{L} = F\dot{x}\dot{\bar{x}} + F\dot{\psi}{\bar{\psi}} + F_{\bar{x}}\dot{\bar{x}}\psi\bar{\psi} - \beta(F\dot{\bar{x}}^2 + F_x\dot{\bar{x}}\bar{\psi}\psi).
\end{equation} 
The superconformal invariance requires that the action of the operator $K$ on $\mathcal{L}$ produces a total derivative. The bosonic and the fermionic parts of the actions can be treated separately and generate the same constraint. As an example, we have for the bosonic sector
\begin{equation*}
K(F\dot{x}\dot{\bar{x}} - \beta F\dot{\bar{x}}^2) = 2t[\lambda F_xx + \lambda F_{\bar{x}}\bar{x} + (1+2\lambda)F](\dot{x}\dot{\bar{x}} + \beta \dot{\bar{x}}^2) + 2\lambda Fx\dot{\bar{x}} + 2\lambda F\bar{x}\dot{x} + 4\beta\lambda F\bar{x}\dot{\bar{x}}.
\end{equation*}
We are led to simultaneously satisfy the system of equations
\begin{eqnarray}
\label{eq12}
\lambda F_xx + \lambda F_{\bar{x}}\bar{x} + (1+2\lambda)F &=&0,\nonumber\\
2\lambda Fx\dot{\bar{x}} + 2\lambda F\bar{x}\dot{x} + 4\beta\lambda F\bar{x}\dot{\bar{x}} &=& \frac{d}{dt}(\ldots ).
\end{eqnarray}
The only solution for the above system is obtained for  $\beta = 0$, with
\begin{equation}
\label{prep1}
F(x,\bar{x}) = C(x\bar{x})^{-\frac{1+2\lambda}{2\lambda}}.
\end{equation}
Therefore, the Lagrangian producing the twisted superconformally invariant $\sigma$-model,
\begin{equation}
\label{invlag1}
\mathcal{L} = F(\dot{x}\dot{\bar{x}} + \dot{\psi}{\bar{\psi}}) + F_{\bar{x}}\dot{\bar{x}}\psi\bar{\psi},
\end{equation}
with $F$ given in (\ref{prep1}), is only recovered at the special critical value $\beta=0$. It is worth pointing out that, for this value,
the $(2,2,0)$ $D$-module rep obtained from  (\ref{qbeta}, \ref{HDpar}, \ref{Kpar}) is reducible.\par
The action derived from (\ref{prep1},\ref{invlag1}) is only defined for $\lambda\neq 0$. It was pointed out in \cite{hoto} that
a consistent $\lambda\rightarrow 0$ limit exists if the propagating bosons are suitably shifted. In the present case this corresponds to replace $x, \bar{x}$ with the shifted fields
$x+ \frac{\rho}{\lambda}, \bar{x} + \frac{\bar{\rho}}{\lambda}$, for $\rho, \bar{\rho}$ constants with the same dimensionality as $x, \bar{x}$. In the $\lambda\rightarrow 0$ limit, a representation of the superconformal algebra which only depends on $\rho, {\bar{\rho}}$ emerges. The corresponding supermultiplet which emerges in this limit has been called the {\em inhomogeneous supermultiplet} in \cite{hoto} (see also \cite{ist}). The construction of the inhomogeneous superconformal actions has been discussed at length in that paper; since it can be straightforwardly applied to the present twisted case, 
it is sufficient to present here the final results. The inhomogeneous invariant superconformal action is expressed by the same Lagrangian (\ref{invlag1}), but with the identification of $F$ given by
\bea\label{prep2}
F(x,\bar{x}) &=& Ce^{-\frac{x + \bar{x}}{2\rho}}.
\eea
The construction of the twisted superconformal action, invariant under the hyperbolic representation, proceeds on similar lines. In the homogeneous $\lambda \neq 0$ case the Lagrangian is
\begin{equation}
\label{invlag2}
\mathcal{L} = F(\dot{x}\dot{\bar{x}} + \mu\dot{\psi}\bar{\psi} + \mu^2\lambda^2 x\bar{x}) + \mu F_{\bar{x}}\psi\bar{\psi}\dot{\bar{x}}, 
\end{equation}
with the same identification of $F$ as in (\ref{prep1}).
In the inhomogeneous $\lambda = 0$ case the invariant Lagrangian is
\begin{equation}
\label{invlag3}
\mathcal{L} = F(\dot{x}\dot{\bar{x}} + \mu\dot{\psi}\bar{\psi} + \mu^2\rho^2) + \mu F_{\bar{x}}\psi\bar{\psi}\dot{\bar{x}},
\end{equation}
with $F$ given by (\ref{prep2}).\par
As it happens for the untwisted superconformal case \cite{pap,hoto}, the difference between parabolic versus hyperbolic superconformal actions consists in the presence, in the latter case, of an extra potential term.  

\subsection{Superconformal models for the $(1,2,1)$ supermultiplet}

The construction of the twisted superconformal actions for the $(1,2,1)$ supermultiplet proceeds along similar lines. A global supersymmetric action is obtained from the Lagrangian  $\mathcal{L} = Q_1Q_2^{(\gamma)}(F\bar{\psi}\psi)$, with $Q_1, Q_2^{(\gamma)}$ given in (\ref{qgamma}). We get
\begin{equation}
\label{eq28}
\mathcal{L} = Fb\dot{x} + F\dot{\psi}\bar{\psi} + \gamma F b^2 + \gamma F_x\bar{\psi}\psi b, 
\end{equation}
where $F = F(x)$.\par
The superconformal invariance is guaranteed if the constraint generated by $K$ is satisfied. In the $(1,2,1)$ case the generator $K$ carries a dependence on a off-diagonal parameter $\delta$, see (\ref{Kdelta}) (for clarity reason, it is therefore convenient to write $K\equiv K^{(\delta)}$). The action of $ K^{(\delta)} $ on the purely bosonic part of the Lagrangian reads as follows
\begin{equation*}
K^{(\delta)}(Fb\dot{x} + \gamma Fb^2) = 2t[\lambda F_x x + (1+2\lambda)F](\gamma b^2 + b\dot{x}) + 2(\lambda + \gamma\delta)Fxb.
\end{equation*}
The superconformal invariance requires simultaneously satisfying the system (no further constraint is obtained from the fermionic part of the Lagrangian)
\bea
\lambda F_x x + (1+2\lambda)F &=& 0,\nonumber\\
\gamma\delta &=& -\lambda.
\eea
The homogeneous case $\lambda\neq 0$ implies that both $\gamma$ and $\delta$ are non-vanishing. This means, in particular, that the $(1,2,1)$ representation of the twisted superconformal algebra entering the invariant action is irreducible.\par  Without loss of generality $\gamma$ can be set to $1$ ($\gamma=1$) via the rescaling of the fields and of the prepotential $F$: $\gamma b \rightarrow b$, $\gamma\bar{\psi} \rightarrow \bar{\psi}$, $\frac{F}{\gamma} \rightarrow F$.
The Lagrangian can indeed be written as
\begin{equation}
\label{eq31}
\mathcal{L} = \frac{1}{\gamma}[F(\gamma b)^2 + F(\gamma b)\dot{x} + F\dot{\psi}(\gamma\bar{\psi}) + F_x(\gamma\bar{\psi})\psi(\gamma b)].
\end{equation}
With the choice $\gamma=1$, $\delta=-\lambda$, the Lagrangian of the parabolic homogeneous case is
\begin{equation}
\label{eq32}
\mathcal{L} = F(b^2 + b\dot x + \dot{\psi}\bar{\psi}) + F_x\bar{\psi}\psi b ,\quad\text{ with } \quad F(x) = Cx^{-\frac{1+2\lambda}{\lambda}}.
\end{equation} 
The parabolic inhomogeneous case is obtained in the $\lambda\rightarrow 0$ limit for operators acting on the shifted supermultiplet $(x+\frac{\rho}{\lambda},b;\psi,\bar{\psi})$. It produces an invariant action whose Lagrangian is
\begin{equation}
\label{eq34}
\mathcal{L} = F(b^2 + b\dot x + \dot{\psi}\bar{\psi}) + F_x\bar{\psi}\psi b, \quad\text{with }\quad F(x) = Ce^{-\frac{x}{\rho}}.
\end{equation}

In the hyperbolic homogeneous case the invariant action is given by the Lagrangian
\begin{equation}
\label{eq37}
\mathcal{L} = F(\mu^2b^2 + \mu b\dot x + \mu\dot{\psi}\bar{\psi}- \mu^2\lambda xb) + F_x(\mu^2\bar{\psi}\psi b+\frac{1}{2}\mu^2\lambda x\psi\bar{\psi}), \quad\text{with } \quad F(x) = Cx^{-\frac{1+2\lambda}{\lambda}}.
\end{equation}
In the hyperbolic inhomogeneous case the invariant action is given by the Lagrangian
\begin{equation}
\label{eq38}
\mathcal{L} = F(\mu^2b^2 + \mu b\dot x + \mu\dot{\psi}\bar{\psi}- \mu^2\rho b) + F_x(\mu^2\bar{\psi}\psi b+\frac{1}{2}\mu^2\rho\psi\bar{\psi}), \quad\text{with }\quad F(x) = Ce^{-\frac{x}{\rho}}.
\end{equation}

\subsection{Superconformal models in the constant kinetic basis}

It is convenient to compare the four inequivalent superconformal actions introduced above by making field redefinitions. Under such redefinitions the superconformal symmetry is realized non-linearly. The Lagrangians, on the other hand, have a simpler form, being recast as a constant kinetic term plus (for $\lambda\neq -\frac{1}{2}$) an interaction.
\par
For the $(2,2,0)$ supermultiplet, in both parabolic and hyperbolic cases, the following field redefinitions are made:\\
{\em i}) in the homogeneous case
\begin{equation}
\label{eq39}
y = -2\lambda x^{-\frac{1}{2\lambda}}, \hspace{1cm} \bar{y} = -2\lambda \bar{x}^{-\frac{1}{2\lambda}}, \hspace{1cm} \chi = \psi, \hspace{1cm}  \bar{\chi} = (x\bar{x})^{-\frac{1+2\lambda}{2\lambda}}\bar{\psi};
\end{equation}
{\em ii}) in the inhomogeneous case
\begin{equation}
\label{eq40}
y = -2\rho e^{-\frac{x}{2\rho}}, \hspace{1cm} \bar{y} = -2\rho e^{-\frac{\bar{x}}{2\rho}},\hspace{1cm} \chi = \psi, \hspace{1cm}  \bar{\chi} = e^{-\frac{x+\bar{x}}{2\rho}}\bar{\psi}.
\end{equation}
For the $(1,2,1)$ supermultiplet, in both parabolic and hyperbolic cases, the field redefinitions are\\
{\em i}) in the homogeneous case
\begin{equation}
\label{eq43}
y = -2\lambda x^{-\frac{1}{2\lambda}},  \hspace{1cm} a = x^{-\frac{1+2\lambda}{2\lambda}} b, \hspace{1cm} \chi = \psi, \hspace{1cm}  \bar{\chi} = x^{-\frac{1+2\lambda}{\lambda}}\bar{\psi};
\end{equation}
{\em ii}) in the inhomogeneous case
\begin{equation}
\label{eq44}
y = -2\rho e^{-\frac{x}{2\rho}}, \hspace{1cm} a = e^{-\frac{x}{2\rho}} b, \hspace{1cm} \chi = \psi, \hspace{1cm}  \bar{\chi} = e^{-\frac{x}{\rho}}\bar{\psi}.
\end{equation}
In terms of the redefined fields the Lagrangians of the superconformally invariant $\sigma$-models are very compactly written. Without loss of generality, in the hyperbolic case the dimensional parameter $\mu $ can be set $\mu=1$ (if needed, we can restore the $\mu$-dependence by dimensional analysis). \par
Associated with the $(2,2,0)$ supermultiplet we have the Lagrangian
\begin{equation}
\label{220act}
\mathcal{L} = \dot{y}\dot{\bar{y}} + \dot{\chi}\bar{\chi} + \frac{\epsilon}{4}y\bar{y}+(1+2\lambda) \frac{\chi\bar{\chi}\dot{\bar{y}}}{\bar{y}}.
\end{equation}

Associated with the $(1,2,1)$ supermultiplet we have the Lagrangian
\begin{equation}
\label{121act}
\mathcal{L} = a^2 +  a\dot{y} + \dot{\chi}\bar{\chi} +\frac{\epsilon}{2}ya + \frac{\epsilon}{2}(1+2\lambda)\bar{\chi}\chi+ 2(1+2\lambda)\frac{\bar{\chi}\chi a}{y}.
\end{equation}
In the above Lagrangians the parameter $\epsilon$ takes the value $\epsilon=0$ in the parabolic case and $\epsilon=1$ in the hyperbolic case. The inhomogeneous case corresponds to the special (non-singular) value
$\lambda=0$. Another special value, $\lambda=-\frac{1}{2}$, corresponds to the non-interacting theory. It is interesting to note that the same twisted superconformal invariance is possessed by actions, even in the presence of an interaction (for $\lambda\neq -\frac{1}{2}$). The coupling constant labels a class of superconformally invariant $\sigma$-models and is expressed as $2\lambda +1$.

\section{SCA $D$-module reps from pseudo-supersymmetry} 
In this Section we summarize the results about inducing $D$-module representations of semi-simple finite superconformal algebras, recovered from pseudo-supersymmetry supermultiplets. This Section extends the results of \cite{kuto, khto, hoto}.
Some of the $D$-module representations below were used to construct the topological conformal algebras discussed in Section {\bf 2}. In the Appendix {\bf B} the $D$-module representations are explicitly given (in terms of operators which allow to reconstruct the complete set of superconformal algebra generators). We have\par
{~}\par
{\bf - The ${\cal N}=(1,1)$-induced superconformal algebra}\par
{~}\par
The superalgebra is $sl(2|1)$. The $D$-module rep is recovered from the $(1,1,0)$ supermultiplet
for any value of the scaling dimension $\lambda$.\par
{~}\par
{\bf - The ${\cal N}=(2,2)$-induced superconformal algebras}\par
{~}\par
The $D$-module reps are recovered from the $(k,2,2-k)$ supermultiplets, with $k=0,1,2$,
for any value of the scaling dimension $\lambda$.\par
The corresponding superconformal algebras are\\
- $D(2,1;\alpha)$, for $(2,2,0)$ and $(0,2,2)$. The relation between $\alpha$ and $\lambda$ is $\alpha = 2(1-k)\lambda$;\\
- $A(1,1)$ for $(1,2,1)$ and at the critical value $\lambda = 0$;\\
- the centrally extended algebra ${\widehat A}(1,1)$ for $(1,2,1)$ and $\lambda \neq 0$.\par
{~}\par
{\bf - The ${\cal N}=(3,3)$-induced superconformal algebras}\par
{~}\par
These superconformal algebras are induced by the supermultiplets $(k,4,4-k)$, for $k\neq 2$, at the critical values of the scaling dimension $\lambda$ given by $\lambda_{cr} = \frac{1}{2(k-2)}$. \par
The identification of the superconformal algebras goes as follows:\\
- $D(3,1)$, recovered from the supermultiplets $(4,4,0)$ and $(0,4,4)$;\\
- $A(2,1)$, recovered from the supermultiplets $(3,4,1)$ and $(1,4,3)$.\par
One should note that in the Kac list of semisimple Lie superalgebras \cite{kac} besides $D(3,1)$ and $A(2,1)$ there exists an extra superalgebra, $B(1,2)$,
which has the property of being one-dimensionally superconformal (see \cite{top2}) with $6+6$ fermionic generators. This superalgebra, however, is not recovered from the 
$(k,4,4-k)$-induced $D$-module reps.{\par}
{~}\par
{\bf - The ${\cal N}=(4,4)$-induced superconformal algebra}\par
{~}\par
The only supermultiplets inducing a $D$-module rep of a superconformal algebra are $(8,8,0)$ and $(0,8,8)$.
The other values of $k$ ($k=1,2,\ldots, 7$) fail to induce a $D$-module rep of a superconformal algebra.
In both cases, the induced superalgebra is $D(4,1)$ and is only recovered at the critical values of the scaling dimension $\lambda$, $\lambda_{cr} = \frac{1}{4}$ for $(8,8,0)$ and  $\lambda_{cr} = -\frac{1}{4}$
for $(0,8,8)$. This case should be compared with the induced $D$-module reps from positive ${\cal N}=(8,0)$ supersymmetry. In that case, see \cite{khto}, the four superconformal algebras $D(4,1)$, $D(2,2)$, $A(3,1)$ 
and $F(4)$ are induced from $D$-module reps of the $(k,8,8-k)$ supermultiplets for different values of $k\neq 4$ at the given critical scaling dimensions. 
\par
{~}\par
{\bf - The ${\cal N}=(5,1)$-induced superconformal algebra}\par
{~}\par
In this case the only supermultiplets inducing a $D$-module rep of a superconformal algebra are $(8,8,0)$ and $(0,8,8)$. The other values of $k$ ($k=1,2,\ldots, 7$) fail to induce a $D$-module rep of a superconformal algebra.
In both cases, the induced superalgebra is $D(3,1)$, recovered at the critical values of the scaling dimension $\lambda$, $\lambda_{cr} = \frac{1}{4}$ for $(8,8,0)$ and  $\lambda_{cr} = -\frac{1}{4}$ for $(0,8,8)$.

\section{Conclusions}

The $sl(2|1)$-invariant topological actions are presented in (\ref{220act}) and (\ref{121act}) for the $(2,2,0)$ chiral and the $(1,2,1)$ real supermultiplet, respectively. The parameter $\epsilon=0,1$ discriminates the parabolic versus the hyperbolic case. A coupling constant is given by $2\lambda +1$; at the $\lambda=-\frac{1}{2}$ value the actions are free.\par
The construction of topological $sl(2|1)$-invariant actions for several supermultiplets in interactions is straightforward. It can be done by adapting to the twisted supersymmetry the construction discussed in \cite{khto} for the untwisted superconformal invariance. \par
At least two approaches can be used to construct topological actions invariant under
a larger superalgebra, let's say $D(2,1;\alpha)$. One can start with $sl(2|1)$-invariant actions from supermultiplets in interactions and then search which values in the space of parameters characterize the enhanced superconformal invariance. To give an example, the $(4,4,0)$ supermultiplet is expected to 
possess $D(2,1;\alpha)$-invariant actions, with $\alpha$ linked to its scaling dimension. Under $sl(2|1)$, the $(4,4,0)$ supermultiplet is decomposed into the direct sum 
of two $(2,2,0)$ supermultiplets: $(4,4,0)= (2,2,0)\oplus (2,2,0)$. \par
The other approach is more direct. It requires the construction of the nilpotent operators (which, together with the boost operator $K$, induce a $D(2,1;\alpha)$ superalgebra) from the linear combination of the $3+3$ pseudo-supersymmetry operators acting (see Appendix {\bf B}) on four bosons and four fermions.
We leave this construction for a future work.\par
In this paper we proved the flexibility of the world-line framework to investigate one-dimensional topological conformal $\sigma$-models \cite{bara}. A natural question, which goes under the name of ``oxidation" \cite{top1} or
``enhancement" \cite{fil2}, is whether data from one-dimensional (twisted) supersymmetry can encode information of higher dimensional theories and whether they allow to reconstruct them. From  the \cite{kawi} paper it is known that two-dimensional topological $\sigma$-models are at the core of the geometric Langlands program. An approach to enhance world-line supersymmetry to world-sheet supersymmetry has been presented in \cite{gahu}. It looks promising investigating its application to topological (i.e. twisted supersymmetric) theories.
{~}
\par
{~}\par{}
\newpage
\renewcommand{\theequation}{A.\arabic{equation}}
\setcounter{equation}{0}
 {~}\\
{\Large{\bf Appendix A: the smallest topological conformal algebra}}\par
{~}~\par
One can easily recognize that the smallest invariant topological conformal algebra is a subalgebra of the ${\cal N}=(1,1)$ superconformal algebra $sl(2|1)$. This invariant algebra, $sl(2){\supset \hspace{-1em}\hspace{-1pt}+}gr(2)$,  is realized on the $(1,1)$ supermultiplet.  It consists of the bosonic $sl(2)$ subalgebra acting on the Grassmann algebra $gr(2)$ (see \cite{fss}) of two anticommuting nilpotent operators $Q, \bar{Q}$. In the parabolic realization the operator $Q$ can be constructed from the positive and negative square roots of $H$, 
\begin{equation}
\label{eq47}
Q^+ = \left(\begin{array}{ccc}
0 & 1 \\
\partial_t & 0 \\
\end{array}\right),
\text{\quad  \quad }
Q^- = \left(\begin{array}{ccc}
0 & 1 \\
-\partial_t & 0 \\
\end{array}\right),
\end{equation}
either as
\begin{equation}
\label{eq48}
\text{{\em i})} \quad Q = \frac{Q^+ + Q^-}{2} = \left(\begin{array}{ccc}
0 & 1 \\
0 & 0 
\end{array}\right)
\text{\hspace{1cm}or\hspace{1cm} }
\text{{\em ii}) }\quad  Q = \frac{Q^+ - Q^-}{2} = \left(\begin{array}{ccc}
0 & 0 \\
\partial_t & 0\end{array}\right).
\end{equation}
The $sl(2)$ generators are
\begin{equation}
\label{eq49}
H = {\mathbb I}_2\cdot\partial_t, \quad D = {\mathbb I}_2\cdot t\partial_t + \Lambda, \quad  K = {\mathbb I}_2\cdot t^2\partial_t + 2t\Lambda, 
\end{equation}
where $\Lambda = diag(\lambda, \lambda + \frac{1}{2})$.
In both cases, the nilpotent partner $\bar{Q}$ is given by $\bar{Q} = [K,Q]$. The invariant action is given by $\mathcal{S}=\int dt\cdot\mathcal{L}$, where the Lagrangian is ${\mathcal L} =  Q(G(x){\dot{{\psi}}})$. In the following it is convenient to introduce
$G(x)=\int^x dx' F(x')$. We have\par
{\em i}) for the first choice,
\begin{equation}
\label{eq50}
\mathcal{L} = F\psi\dot{\psi},
\end{equation}
where for the {\em homogeneous/inhomogeneous} cases discussed in Section {\bf 3} we have, respectively,  $F(x) = Cx^{-\frac{1+2\lambda}{\lambda}}$ and $F(x) = Ce^{-\frac{x}{\rho}}$;\par
{\em ii}) for the second choice,
\begin{equation}
\label{eq51}
\mathcal{L} = F\dot{x}^2 + V(x),
\end{equation}
with $F(x) = Cx^{-\frac{1+2\lambda}{\lambda}}$, $V(x) = kx^{\frac{1}{\lambda}}$ in the 
{homogeneous case}  and $F(x) = Ce^{-\frac{x}{\rho}}$, $V(x) = ke^{\frac{x}{\rho}}$  in the {inhomogeneous case} .\par
In the second case the potential term $V(x)$ is automatically invariant, up to a time derivative, so that the
parameter $k$ is arbitrary.  \par
The construction can be repeated for the hyperbolic realization. We obtain in this case
\begin{equation}
\label{eq52}
{\text {\em i}) }\quad
Q = e^{-\frac{\mu t}{2}}\left(\begin{array}{ccc}
0 & 1 \\
0 & 0 \\
\end{array}\right)
\text{\hspace{1cm} or \hspace{1cm} }
\text{{\em ii}) }\quad
Q = e^{-\frac{\mu t}{2}}\left(\begin{array}{ccc}
0 & 0 \\
\frac{1}{\mu}\partial_t - \lambda & 0 \\
\end{array}\right).
\end{equation}
The invariant actions are given by the Lagrangians\par
{\em i}) for the first choice,
\begin{equation}
\label{eq53}
\mathcal{L} = \mu F\psi\dot{\psi}\text{, }
\end{equation}
with $F(x) = Cx^{-\frac{1+2\lambda}{\lambda}}$ in the {homogeneous case}   and $F(x) = Ce^{-\frac{x}{\rho}}$ in the {inhomogeneous case};\par
{\em ii}) for the second choice,
\begin{equation}
\label{eq54}
\mathcal{L} = F[\dot{x}^2 + \mu^2(\lambda x + \rho)^2] + V(x)\text{, }
\end{equation}
with $F(x) = Cx^{-\frac{1+2\lambda}{\lambda}}$, $V(x) = kx^{\frac{1}{\lambda}}$ and $\rho$ = 0 in the {homogeneous case} and $F(x) = Ce^{-\frac{x}{\rho}}$, $V(x) = ke^{\frac{x}{\rho}}$ and $\lambda = 0$  in the {inhomogeneous case}.

\par
{~}\par
\renewcommand{\theequation}{B.\arabic{equation}}
\setcounter{equation}{0}
 
{\Large{\bf Appendix B: explicit $D$-module reps of the topological superconformal algebras}}\par
~\par

We explicitly give here, for each representation discussed in Section {\bf 4}, a set of operators which allows to recover (via repeated mutual commutators/anticommutators) the complete set of superconformal algebra generators. It is sufficient to present the parabolic case. In the following ${\mathbb I}_n$ denotes the $n\times n$ identity matrix, while $E_{ij}$ stands for the matrix whose only non-vanishing entry is $1$ at the crossing of the $i$-th row with the $j$-th column.\par
For all the $D$-module representations below the generators $H,D$ of the $sl(2)$ Borel subalgebra have the form
\bea
H&=&{\mathbb I}_{n}\cdot\partial_t,\nonumber\\
D&=&{\mathbb I}_{n}\cdot t\partial_t + \Lambda,
\eea
for a given integer $n$ and a given diagonal matrix $\Lambda$.\par
One can prove that the remaining $sl(2)$ generator, $K$, is in diagonal form for the $(1,1,0)$ $D$-module reps of the ${\cal N}=(1,1)$ superconformal algebra and the $(2,2,0)$, $(0,2,2)$ $D$-module reps of the ${\cal N}=(2,2)$ superconformal algebras. $K$ admits a consistent non-diagonal form (parametrized by the free, off-diagonal parameter $\delta$, the diagonal case being recovered at $\delta=0$) for the $(1,2,1)$ $D$-module reps of the ${\cal N}=(2,2)$ superconformal algebras. In this case $K$ is given by formula (\ref{Kdelta}),
that is $
K = {\mathbb I}_4\cdot t^2\partial_t + 2\Lambda t + \delta E_{21},
$
with $\Lambda = diag(\lambda, \lambda +1, \lambda+\frac{1}{2}, \lambda + \frac{1}{2})$.\par
In all remaining cases with extended number of supersymmetries, ${\cal N}=(3,3)$, ${\cal N}=(4,4)$ and ${\cal N}=(5,1)$,
 we investigated the closure of the representations for diagonal $K$.
Therefore, in the following, apart the $(1,2,1)$ $D$-module rep of the ${\cal N}=(2,2)$ superconformal algebra
with $K$ given by (\ref{Kdelta}), we have, in all other cases,
\bea
K&=&{\mathbb I}_{n}\cdot t^2\partial_t + 2\Lambda t.
\eea

\subsubsection*{The ${\cal N}=(1,1)$ superconformal algebra}

In this case the unique supermultiplet is $(1,1,0)$, with component fields $(x;\psi)$.\\
We have $n=2$, while the matrix $\Lambda$ is given, for an arbitrary $\lambda$, by $\Lambda = diag(\lambda, \lambda+\frac{1}{2})$. \\
The square roots of $\pm H$ are the operators\\
\noindent\(Q^+= E_{12} + E_{21}\partial_t\),\\
\noindent\(Q^-= E_{12} - E_{21}\partial_t\).\\

\subsection*{The ${\cal N}=(2,2)$ superconformal algebras}

In this case $n=4$. Different superconformal algebras, see Section {\bf 4}, are associated with different values of the scaling dimension $\lambda$.
The representations are labeled by the supermultiplets\\
\newline {\bf -  $ (2,2,0)$:}\\
Here $\Lambda = diag(\lambda, \lambda, \lambda+\frac{1}{2}, \lambda + \frac{1}{2})$. The component fields are $(x,\bar{x};\psi,\bar{\psi})$, where $x$, $\bar{x}$ are propagating bosons.  \newline
The square roots of $\pm H$ are the operators\\
\noindent\(Q_1^+= E_{14} + E_{23} + (E_{32} + E_{41})\partial_t,\)\\
\noindent\(Q_2^+= E_{13} - E_{24} + (E_{31} - E_{42})\partial_t,\)\\
\noindent\(Q_1^-= E_{14} - E_{23} + (E_{32} - E_{41})\partial_t,\)\\
\noindent\(Q_2^-= E_{13} + E_{24}  - (E_{31} + E_{42})\partial_t.\)\\
\newline
{\bf -  $ (1,2,1)$:}\\
Here $\Lambda = diag(\lambda, \lambda+1 , \lambda+\frac{1}{2}, \lambda + \frac{1}{2})$. The component fields are $(x,b;\psi,\bar{\psi})$, where $x$ is the propagating boson.  \newline
The square roots of $\pm H$ are the operators\\
\noindent\(Q_1^+= E_{14} + E_{32} + (E_{23} + E_{41})\partial_t,\)\\
\noindent\(Q_2^+= E_{13} - E_{42} + (-E_{24} + E_{31})\partial_t, \)\\
\noindent\(Q_1^-= E_{14} + E_{32} - (E_{23} + E_{41})\partial_t,\)\\
\noindent\(Q_2^-= E_{13} - E_{42} + (E_{24} - E_{31})\partial_t. \)\\
\newline
{\bf -  $ (0,2,2)$:}\\
Here $\Lambda = diag(\lambda+1, \lambda+1 , \lambda+\frac{1}{2}, \lambda + \frac{1}{2})$.
This supermultiplet has no propagating bosons. The component fields are $(b, \bar{b};\psi,\bar{\psi})$.\newline
The square roots of $\pm H$ are the operators\\
\noindent\(Q_1^+= (E_{14} + E_{23})\partial_t + E_{32} + E_{41},\)\\
\noindent\(Q_2^+= (E_{13} - E_{24})\partial_t + E_{31} - E_{42},\)\\
\noindent\(Q_1^-= (E_{14} - E_{23})\partial_t + E_{32} - E_{41},\)\\
\noindent\(Q_2^-= (E_{13} + E_{24})\partial_t - E_{31} - E_{42}.\)
\newline

\subsection*{The ${\cal N}=(3,3)$ superconformal algebras}
In this case $n=8$. Different superconformal algebras, see Section {\bf 4}, are associated with different critical values of the scaling dimension $\lambda$.
The representations are labeled by the supermultiplets\\
\newline {\bf -  $ (4,4,0)$:}\\
This superalgebra is critical at $\lambda = \frac{1}{4}$. We have $\Lambda = diag(\lambda, \lambda, \lambda, \lambda, \lambda+\frac{1}{2}, \lambda + \frac{1}{2},\lambda+\frac{1}{2},\lambda+\frac{1}{2})$.\newline
The square roots of $\pm H$ are the operators\\
\noindent\(Q_1^+=E_{18}-E_{27}-E_{36}+E_{45}+ (E_{54} - E_{63} - E_{72} + E_{81})\partial_t,\)\\
\noindent\(Q_2^+=E_{17}+E_{28}+E_{35}+E_{46}+ (E_{53} + E_{64} + E_{71} + E_{82})\partial_t,\)\\
\noindent\(Q_3^+=E_{15}+E_{26}-E_{37}-E_{48}+ (E_{51} + E_{62} - E_{73} - E_{84})\partial_t,\)\\
\noindent\(Q_1^-=E_{18}+E_{27}-E_{36}-E_{45}+ (E_{54} + E_{63} - E_{72} - E_{81})\partial_t,\)\\
\noindent\(Q_2^-=E_{17}-E_{28}-E_{35}+E_{46}+ (E_{53} - E_{64} - E_{71} + E_{82})\partial_t,\)\\
\noindent\(Q_3^-=E_{15}+E_{26}+E_{37}+E_{48}- (E_{51} + E_{62} + E_{73} + E_{84})\partial_t.\)\newline
\newline
{\bf -  $ (3,4,1)$:}\\
This superalgebra is critical at $\lambda = \frac{1}{2}$. We have $\Lambda = diag(\lambda, \lambda, \lambda, \lambda+1, \lambda+\frac{1}{2}, \lambda + \frac{1}{2},\lambda+\frac{1}{2},\lambda+\frac{1}{2})$.\newline
The square roots of $\pm H$ are the operators\\
\noindent\(Q_1^+=E_{18}-E_{27}-E_{36}+ E_{54} +(E_{45} - E_{63} - E_{72} + E_{81})\partial_t,\)\\
\noindent\(Q_2^+=E_{17}+E_{28}+E_{35}+ E_{64} +(E_{46}+ E_{53} + E_{71}+ E_{82})\partial_t,\)\\
\noindent\(Q_3^+=E_{15}+E_{26}-E_{37}- E_{84} +(-E_{48}+ E_{51}+ E_{62}- E_{73})\partial_t,\)\\
\noindent\(Q_1^-=E_{18}+E_{27}-E_{36}+ E_{54}+(-E_{45}+ E_{63}- E_{72}- E_{81})\partial_t,\)\\
\noindent\(Q_2^-=E_{17}-E_{28}-E_{35}- E_{64}+(E_{46}+ E_{53}- E_{71}+ E_{82})\partial_t,\)\\
\noindent\(Q_3^-=E_{15}+E_{26}+E_{37}- E_{84}+(E_{48}- E_{51}- E_{62}- E_{73})\partial_t.\)\\
\newline
{\bf -  $ (1,4,3)$:}\\
This superalgebra is critical at $\lambda = -\frac{1}{2} $. We have $\Lambda = diag(\lambda, \lambda+1, \lambda+1, \lambda+1, \lambda+\frac{1}{2}, \lambda + \frac{1}{2},\lambda+\frac{1}{2},\lambda+\frac{1}{2})$.\newline
The square roots of $\pm H$ are the operators\\
\noindent\(Q_1^+=(-E_{27}-E_{36}+E_{45}+ E_{81})\partial_t+ E_{18}+ E_{54}- E_{63}- E_{72},\)\\
\noindent\(Q_2^+=(E_{28}+E_{35}+E_{46}+ E_{71})\partial_t+ E_{17}+E_{53}+ E_{64}+ E_{82},\)\\
\noindent\(Q_3^+=(E_{26}-E_{37}-E_{48}+ E_{51})\partial_t+ E_{15}+E_{62}- E_{73}- E_{84},\)\\
\noindent\(Q_1^-=(E_{27}-E_{36}-E_{45}- E_{81})\partial_t+E_{18}+ E_{54}+ E_{63}- E_{72},\)\\
\noindent\(Q_2^-=(-E_{28} -E_{35} +E_{46} - E_{71})\partial_t+ E_{17}+ E_{53}- E_{64}+ E_{82},\)\\
\noindent\(Q_3^-=(E_{26}+E_{37}+E_{48}- E_{51})\partial_t +E_{15}- E_{62}- E_{73}- E_{84}.\)\\
\newline
{\bf -  $ (0,4,4)$:}\\
The superalgebra is critical at $\lambda = -\frac{1}{4}$. We have $\Lambda = diag(\lambda+1, \lambda+1, \lambda+1, \lambda+1, \lambda+\frac{1}{2}, \lambda + \frac{1}{2},\lambda+\frac{1}{2},\lambda+\frac{1}{2})$.\newline
The square roots of $\pm H$ are the operators\\
\noindent\(Q_1^+=(E_{18}-E_{27}-E_{36}+E_{45})\partial_t+ E_{54}- E_{63}- E_{72}+ E_{81},\)\\
\noindent\(Q_2^+=(E_{17}+E_{28}+E_{35}+E_{46})\partial_t+ E_{53}+ E_{64}+ E_{71}+ E_{82},\)\\
\noindent\(Q_3^+=(E_{15}+E_{26}-E_{37}-E_{48})\partial_t+ E_{51}+ E_{62}- E_{73}- E_{84},\)\\
\noindent\(Q_1^-=(E_{18}+E_{27}-E_{36}-E_{45})\partial_t+ E_{54}+ E_{63}- E_{72}- E_{81},\)\\
\noindent\(Q_2^-=(E_{17}-E_{28}-E_{35}+E_{46})\partial_t+ E_{53}- E_{64}- E_{71}+ E_{82},\)\\
\noindent\(Q_3^-=(E_{15}+E_{26}+E_{37}+E_{48})\partial_t- E_{51}- E_{62}- E_{73}- E_{84}.\)
\newline

\subsection*{The ${\cal N}=(4,4)$ superconformal algebra}
In this case $n=16$. There is a unique superconformal algebra, see Section {\bf 4}. Its two representations are given by the supermultiplets\\
\newline {\bf -  $ (8,8,0)$:}\\
We have $\Lambda = diag(\lambda,\lambda,\lambda,\lambda,\lambda,\lambda,\lambda,\lambda,\lambda + \frac{1}{2},\lambda+\frac{1}{2},\lambda+\frac{1}{2},\lambda+\frac{1}{2},\lambda+\frac{1}{2},\lambda+\frac{1}{2},\lambda+\frac{1}{2},\lambda+\frac{1}{2})$, at the critical value, due to the closure conditions,  $\lambda = \frac{1}{4}$. \newline
The square roots of $\pm H$ are the operators\\
\noindent\(Q_1^+=E_{1,16}+E_{2,15}+E_{3,14}+E_{4,13}+E_{5,12}+E_{6,11}+E_{7,10}+E_{8,9}+(E_{9,8}+ E_{10,7}+
E_{11,6}+ E_{12,5}+ E_{13,4}+ E_{14,3}+ E_{15,2}+ E_{16,1})\partial_t,\)\\
\noindent\(Q_2^+=E_{1,15}-E_{2,16}+E_{3,13}-E_{4,14}+E_{5,11}-E_{6,12}+E_{7,9}-E_{8,10}+(E_{9,7}- E_{10,8}+
E_{11,5}- E_{12,6}+ E_{13,3}- E_{14,4}+ E_{15,1}- E_{16,2})\partial_t,\)\\
\noindent\(Q_3^+=E_{1,13}+E_{2,14}-E_{3,15}-E_{4,16}+E_{5,9}+E_{6,10}-E_{7,11}-E_{8,12}+ ( E_{9,5}+ E_{10,6}-
E_{11,7}- E_{12,8}+ E_{13,1}+ E_{14,2}- E_{15,3}- E_{16,4})\partial_t,\)\\
\noindent\(Q_4^+=E_{1,9}+E_{2,10}+E_{3,11}+E_{4,12}-E_{5,13}-E_{6,14}-E_{7,15}-E_{8,16}+ ( E_{9,1}+ E_{10,2}+
E_{11,3}+ E_{12,4}- E_{13,5}- E_{14,6}- E_{15,7}- E_{16,8})\partial_t,\)\\
\noindent\(Q_1^-=E_{1,16}-E_{2,15}+E_{3,14}-E_{4,13}+E_{5,12}-E_{6,11}+E_{7,10}-E_{8,9}+ ( E_{9,8}- E_{10,7}+
E_{11,6}- E_{12,5}+ E_{13,4}- E_{14,3}+ E_{15,2}- E_{16,1})\partial_t,\)\\
\noindent\(Q_2^-=E_{1,15}+E_{2,16}-E_{3,13}-E_{4,14}+E_{5,11}+E_{6,12}-E_{7,9}-E_{8,10}+  (E_{9,7}+ E_{10,8}-
E_{11,5}- E_{12,6}+ E_{13,3}+ E_{14,4}- E_{15,1}- E_{16,2})\partial_t,\)\\
\noindent\(Q_3^-=E_{1,13}+E_{2,14}+E_{3,15}+E_{4,16}-E_{5,9}-E_{6,10}-E_{7,11}-E_{8,12}+   (E_{9,5}+ E_{10,6}+
E_{11,7}+ E_{12,8}- E_{13,1}- E_{14,2}- E_{15,3}- E_{16,4})\partial_t,\)\\
\noindent\(Q_4^-=E_{1,9}+E_{2,10}+E_{3,11}+E_{4,12}+E_{5,13}+E_{6,14}+E_{7,15}+E_{8,16}-  (E_{9,1}+E_{10,2}+
E_{11,3}+ E_{12,4}+ E_{13,5}+ E_{14,6}+ E_{15,7}+ E_{16,8})\partial_t.\)\\
\newline
\bf -  $ (0,8,8)$:}\\
We have $\Lambda = diag(\lambda+1,\lambda+1,\lambda+1,\lambda+1,\lambda+1,\lambda+1,\lambda+1,\lambda+1,\lambda + \frac{1}{2},\lambda+\frac{1}{2},\lambda+\frac{1}{2},\lambda+\frac{1}{2},\lambda+\frac{1}{2},\lambda+\frac{1}{2},\lambda+\frac{1}{2},\lambda+\frac{1}{2})$ at the critical value $\lambda = -\frac{1}{4}$. \newline
The square roots of $\pm H$ are the operators\\
\noindent\(Q_1^+=(E_{1,16}+E_{2,15}+E_{3,14}+E_{4,13}+E_{5,12}+E_{6,11}+E_{7,10}+E_{8,9})\partial_t +E_{9,8}+ E_{10,7}+
E_{11,6}+ E_{12,5}+ E_{13,4}+ E_{14,3}+ E_{15,2}+ E_{16,1},\)\\
\noindent\(Q_2^+=(E_{1,15}-E_{2,16}+E_{3,13}-E_{4,14}+E_{5,11}-E_{6,12}+E_{7,9}-E_{8,10})\partial_t +E_{9,7}- E_{10,8}+
E_{11,5}- E_{12,6}+ E_{13,3}- E_{14,4}+ E_{15,1}- E_{16,2},\)\\
\noindent\(Q_3^+=(E_{1,13}+E_{2,14}-E_{3,15}-E_{4,16}+E_{5,9}+E_{6,10}-E_{7,11}-E_{8,12})\partial_t +  E_{9,5}+ E_{10,6}-
E_{11,7}- E_{12,8}+ E_{13,1}+ E_{14,2}- E_{15,3}- E_{16,4},\)\\
\noindent\(Q_4^+=(E_{1,9}+E_{2,10}+E_{3,11}+E_{4,12}-E_{5,13}-E_{6,14}-E_{7,15}-E_{8,16})\partial_t +  E_{9,1}+ E_{10,2}+
E_{11,3}+ E_{12,4}- E_{13,5}- E_{14,6}- E_{15,7}- E_{16,8},\)\\
\noindent\(Q_1^-=(E_{1,16}-E_{2,15}+E_{3,14}-E_{4,13}+E_{5,12}-E_{6,11}+E_{7,10}-E_{8,9})\partial_t +  E_{9,8}- E_{10,7}+
E_{11,6}- E_{12,5}+ E_{13,4}- E_{14,3}+ E_{15,2}- E_{16,1},\)\\
\noindent\(Q_2^-=(E_{1,15}+E_{2,16}-E_{3,13}-E_{4,14}+E_{5,11}+E_{6,12}-E_{7,9}-E_{8,10})\partial_t +  E_{9,7}+ E_{10,8}-
E_{11,5}- E_{12,6}+ E_{13,3}+ E_{14,4}- E_{15,1}- E_{16,2},\)\\
\noindent\(Q_3^-=(E_{1,13}+E_{2,14}+E_{3,15}+E_{4,16}-E_{5,9}-E_{6,10}-E_{7,11}-E_{8,12})\partial_t +   E_{9,5}+ E_{10,6}+
E_{11,7}+ E_{12,8}- E_{13,1}- E_{14,2}- E_{15,3}- E_{16,4},\)\\
\noindent\(Q_4^-=(E_{1,9}+E_{2,10}+E_{3,11}+E_{4,12}+E_{5,13}+E_{6,14}+E_{7,15}+E_{8,16})\partial_t -  (E_{9,1}+E_{10,2}+
E_{11,3}+ E_{12,4}+ E_{13,5}+ E_{14,6}+ E_{15,7}+ E_{16,8}).\)\\

\subsection*{The ${\cal N}=(5,1)$ superconformal algebra}
In this case $n=16$. There is a unique superconformal algebra, see Section {\bf 4}. Its two representations are given by the supermultiplets\\
\newline {\bf -  $ (8,8,0)$:}\\
We have $\Lambda = diag(\lambda,\lambda,\lambda,\lambda,\lambda,\lambda,\lambda,\lambda,\lambda + \frac{1}{2},\lambda+\frac{1}{2},\lambda+\frac{1}{2},\lambda+\frac{1}{2},\lambda+\frac{1}{2},\lambda+\frac{1}{2},\lambda+\frac{1}{2},\lambda+\frac{1}{2})$ at the critical value $\lambda = \frac{1}{4}$. \newline
The square roots of $\pm H$ are the operators\\
\noindent\(Q_1^+=E_{1,9}+E_{2,10}+E_{3,11}+E_{4,12}-E_{5,13}-E_{6,14}-E_{7,15}-E_{8,16} + (E_{9,1}+ E_{10,2}+
E_{11,3}+ E_{12,4}- E_{13,5}- E_{14,6}- E_{15,7}- E_{16,8})\partial_t,\)\\
\noindent\(Q_2^+=E_{1,16}-E_{2,15}+E_{3,14}-E_{4,13}-E_{5,12}+E_{6,11}-E_{7,10}+E_{8,9} +  (E_{9,8}- E_{10,7}+
E_{11,6}- E_{12,5}- E_{13,4}+ E_{14,3}- E_{15,2}+ E_{16,1})\partial_t,\)\\
\noindent\(Q_3^+=E_{1,15}+E_{2,16}-E_{3,13}-E_{4,14}-E_{5,11}-E_{6,12}+E_{7,9}+E_{8,10} + ( E_{9,7}+ E_{10,8}-
E_{11,5}- E_{12,6}- E_{13,3}- E_{14,4}+ E_{15,1}+ E_{16,2})\partial_t,\)\\
\noindent\(Q_4^+=E_{1,14}-E_{2,13}-E_{3,16}+E_{4,15}-E_{5,10}+E_{6,9}+E_{7,12}-E_{8,11} + ( E_{9,6}- E_{10,5}-
E_{11,8}+ E_{12,7}- E_{13,2}+ E_{14,1}+ E_{15,4}- E_{16,3})\partial_t,\)\\
\noindent\(Q_5^+=E_{1,13}+E_{2,14}+E_{3,15}+E_{4,16}+E_{5,9}+E_{6,10}+E_{7,11}+E_{8,12} + ( E_{9,5}+ E_{10,6}+
E_{11,7}+ E_{12,8}+ E_{13,1}+ E_{14,2}+ E_{15,3}+ E_{16,4})\partial_t,\)\\
\noindent\(Q_1^-=E_{1,9}+E_{2,10}+E_{3,11}+E_{4,12}+E_{5,13}+E_{6,14}+E_{7,15}+E_{8,16} - ( E_{9,1}+ E_{10,2}+
E_{11,3}+ E_{12,4}+ E_{13,5}+ E_{14,6}+ E_{15,7}+ E_{16,8})\partial_t.\)\\
\newline {\bf -  $ (0,8,8)$:}\\
We have $\Lambda = diag(\lambda+1,\lambda+1,\lambda+1,\lambda+1,\lambda+1,\lambda+1,\lambda+1,\lambda+1,\lambda + \frac{1}{2},\lambda+\frac{1}{2},\lambda+\frac{1}{2},\lambda+\frac{1}{2},\lambda+\frac{1}{2},\lambda+\frac{1}{2},\lambda+\frac{1}{2},\lambda+\frac{1}{2})$ at the critical value $\lambda = -\frac{1}{4}$. \newline
The square roots of $\pm H$ are the operators\\
\noindent\(Q_1^+=(E_{1,9}+E_{2,10}+E_{3,11}+E_{4,12}-E_{5,13}-E_{6,14}-E_{7,15}-E_{8,16})\partial_t + E_{9,1}+ E_{10,2}+
E_{11,3}+ E_{12,4}- E_{13,5}- E_{14,6}- E_{15,7}- E_{16,8},\)\\
\noindent\(Q_2^+=(E_{1,16}-E_{2,15}+E_{3,14}-E_{4,13}-E_{5,12}+E_{6,11}-E_{7,10}+E_{8,9})\partial_t +  E_{9,8}- E_{10,7}+
E_{11,6}- E_{12,5}- E_{13,4}+ E_{14,3}- E_{15,2}+ E_{16,1},\)\\
\noindent\(Q_3^+=(E_{1,15}+E_{2,16}-E_{3,13}-E_{4,14}-E_{5,11}-E_{6,12}+E_{7,9}+E_{8,10})\partial_t +  E_{9,7}+ E_{10,8}-
E_{11,5}- E_{12,6}- E_{13,3}- E_{14,4}+ E_{15,1}+ E_{16,2},\)\\
\noindent\(Q_4^+=(E_{1,14}-E_{2,13}-E_{3,16}+E_{4,15}-E_{5,10}+E_{6,9}+E_{7,12}-E_{8,11})\partial_t +  E_{9,6}- E_{10,5}-
E_{11,8}+ E_{12,7}- E_{13,2}+ E_{14,1}+ E_{15,4}- E_{16,3},\)\\
\noindent\(Q_5^+=(E_{1,13}+E_{2,14}+E_{3,15}+E_{4,16}+E_{5,9}+E_{6,10}+E_{7,11}+E_{8,12})\partial_t +  E_{9,5}+ E_{10,6}+
E_{11,7}+ E_{12,8}+ E_{13,1}+ E_{14,2}+ E_{15,3}+ E_{16,4},\)\\
\noindent\(Q_1^-=(E_{1,9}+E_{2,10}+E_{3,11}+E_{4,12}+E_{5,13}+E_{6,14}+E_{7,15}+E_{8,16})\partial_t - ( E_{9,1}+ E_{10,2}+
E_{11,3}+ E_{12,4}+ E_{13,5}+ E_{14,6}+ E_{15,7}+ E_{16,8}).\)\\
{~}
\\ {~}~
\\ {\Large{\bf Acknowledgments}}
{}~\par{}~\par
We are grateful to S. Khodaee and Z. Kuznetsova for discussions.
This research was supported by CNPq under PQ Grant 306333/2013-9 and FAPERJ under Grant E-26/101.538/2014.
L. B. is grateful to CBPF for hospitality under a PCI-BEV grant and to EPLANET for partial support.

\end{document}